\documentstyle[preprint,aps]{revtex}

\tighten

\begin{document}
\title{{\bf Numerical study of the density of states for the bag model}}
\author{Y Zhang$^{*}\P $, W L Qian$^{*}\P $, S Q Ying$\P $ and R K Su$\S \P $}
\address{$^{*}$Surface Physics Lab.(National Key Lab.), Fudan University, Shanghai\\
200433, P.R.China\\
$\P $Physics Department, Fudan University, Shanghai 200433, P.R.China\\
$\S $CCAST(World Laboratory), P.O.Box 8730, Beijing 100080, P.R.China}
\maketitle

\begin{abstract}
The density of states for an extended MIT bag model is studied numerically
by using a parameterized smooth representation which provides the best fit
to the numerical data. It is found that the mass dependence of the surface
term in the density of states agrees with that derived from multi-reflection
theory calculation. The mass dependence of the curvature term in the density
of states is extracted for finite values of the quark mass. The scaling
properties of the data, which is assumed in the parameterization, are
studied. The difference between the parameterized smooth representation
determined by a best fit and the results derived from direct smoothing of
the numerical data are discussed. The fluctuations in the density of states
are also discussed. We provide a smooth representation of the density of
states of a spherical cavity for non-relativistic particles.

PACS number(s): 12.39.Ba, 12.39.Ki, 21.10.Ma and 12.39.x.
\end{abstract}

\vspace{1cm}

\section{Introduction}

Since the bag model suggested by the MIT group in the 1970s, many other bag
models including the chiral bag \cite{1}, soliton bag \cite{2} and
quark-meson coupling model \cite{3} which take the quark and meson coupling
interactions into account, have been put forward with widely successful
applications in nuclear physics. Recently, the bag model has been extended
to include strange quarks \cite{3-1,4}, and has been employed to study the
properties of a strangelet \cite{5}-\cite{7} which is believed, if it
exists, to be an unambiguous signature for the formation of quark-gluon
plasma \cite{8}.

To calculate the dynamical and/or thermodynamical quantities of spherical
nuclei or a spherical droplet of hadronic or quark matter, in particular the
strange quark matter, by means of the bag model, we must first look for the
density of states (DOS) of the spherical system. The DOS of a spherical
cavity in which the free particles are contained can be expressed as \cite{9}
\begin{equation}
\rho (k)={\frac{dN(k)}{dk}},  \label{a}
\end{equation}
where $N(k)$ is the total number of particle states. In general, the quantum
states are discrete and $N(k)$ is a summation of discontinuous states. But
in practical calculation, a continuous representation of the DOS is needed
for changing the summation to an integral. It will bring some errors and we
will discuss the problem in detail below.

Usually, $N(k)$ can be written in terms of the dimensionless variable $kR$
as 
\begin{equation}
N(k)=A(kR)^3+B(kR)^2+C(kR)+N_{res}^{(0)}(k),  \label{totNexp1}
\end{equation}
where $R$ is the radius of the bag. The first, second and third terms of the
right hand side of Eq.(\ref{totNexp1}) refer to the contributions of the
volume, surface and curvature, respectively, and $N_{res}^{(0)}(k)$ refers
to other contributions which are not proportional to $(kR)^3$, $(kR)^2$ or $%
(kR)$. We will discuss the term $N_{res}^{(0)}(k)$ in section 3 in detail.
In general, the coefficients $A$, $B$ and $C$ are expected to be functions
of $m/k$, and their expressions are model dependent. For example, using the
box renormalization of the wave function, Jaqamin et al obtained \cite{10}

\begin{equation}
A=\frac 2{9\pi },B=-\frac 14,C=\frac 14.  \label{3}
\end{equation}
In their approximation, $A$, $B$ and $C$ are all constants and do not depend
on the momentum. Of course, this treatment seems to be too rough because the
plane wave is not the real eigenmode of the spherical bag geometry. The
exact eigenfunctions of the spherical bag model, as shown earlier, are the
spherical Bessel functions.

Another approach which is better than the plane wave approximation is the
multi-reflection theory (MRT) \cite{11},\cite{12}. According to MRT, the
volume term $A$ can be fixed using the known total number of states for an
infinite system with very large $R$, 
\begin{equation}
\lim_{R\to \infty }A=A_\infty =\frac{2g}{9\pi },  \label{4}
\end{equation}
where $g$ is the total number of degeneracies. For example, $g$ is the total
number of spin and color degrees of freedom for a quark with flavor treated
separately. The surface term $B$ was approximately derived from MRT to be 
\cite{5} 
\begin{equation}
B_{MRT}\left( \frac mk\right) =\frac g{2\pi }\left\{ \left[ 1+\left( \frac mk%
\right) ^2\right] \tan ^{-1}\left( \frac km\right) -\left( \frac mk\right) -%
\frac \pi 2\right\} ,  \label{5}
\end{equation}
where $m$ is the mass of a quark. The curvature term $C$ has not even been
evaluated the present except for the two limiting cases: $m\to 0$ and $m\to
\infty $ according to MRT. Madsen proposed \cite{13},\cite{14} an arbitrary
form for the curvature term $C$ that simply connects the above mentioned
limits, namely 
\begin{equation}
\widetilde{C}({\frac mk})={\frac g{2\pi }}\left\{ \frac 13+\left( {\frac km}+%
{\frac mk}\right) \tan ^{-1}{\frac km}-{\frac{\pi k}{2m}}\right\} ,
\label{6}
\end{equation}
to study the strangelet. Of course, this is only an ansatz with two correct
limits.

Since the spherical bag geometry is popular for finite nuclei or quark
systems and the DOS plays a key role in calculations, it is of interest to
study this problem by exact numerical calculations under such a geometry.
Gilson and Jaffe \cite{15} evaluated the microscopic total number of quark
states $N(k)$ numerically under the MIT bag boundary condition for a fixed
quark mass $m=150MeV$. This is appropriate to the nonrelativistic limit $m>k$
. To our knowledge, the mass dependence of the DOS found by numerical
calculation is still lacking. The purpose of this paper is to check the MRT,
especially the surface term given by Eq.(\ref{5}) and the curvature term by
Eq.(\ref{6}), and to find the mass dependence of the DOS by numerical
counting of the eigenmode under the MIT bag boundary condition. This is of
value because our knowledge of $N(k)$ is improved through the following
analysis:

(1) A numerical study of the mass dependence of the surface term and the
curvature term using a smooth representation which would best fit the total
number of quark states obtained from the numerical data. (2) Comparison of
the surface term to the MRT result to provide a first principle's check of
the theory. (3) Extraction of the unknown curvature term from the numerical
data to provide a suitable smooth representation, instead of Eq.(\ref{6}),
our representation is not only correct at the two limits $m\to 0$ and $m\to
\infty $, but can also be used in the whole mass region. (4) Study of the
extent to which the smooth representation of $N(k)$ can be applied.

The organization of this paper is as follows: Section 2 contains a
description of the procedure for numerical calculation of the total number
of quark states. In section 3, the resulting fluctuating numerical data are
fitted by using a smooth representation for the same quantity to obtain the
coefficients of the volume, surface and curvature terms. The fluctuations
and their behaviour under smoothing of the remaining part of the numerical
data are also studied in this section. Section 4 is our conclusion,
discussion and the results for the DOS using a non-relativistic boundary
condition.

\section{Numerical Evaluation}

The quarks inside a MIT bag are free Dirac particles bouncing back and forth
from the infinite barrier on the bag surface. Their discrete energies in a
spherical bag are determined by the MIT boundary condition \cite{16} 
\begin{equation}
\left. \left( in_\mu \gamma ^\mu \psi -\psi \right) \right| _{\partial V}=0,
\label{7}
\end{equation}
where $\psi $ is the wave function for quarks, $\partial V$ the MIT bag
surface and $n^\mu $ the outward unit vector\footnote{%
The convention for the metric tensor adopted here is $\{g^{\mu\nu}\}= %
\mbox{diag}\{1,-1,-1,-1\}$.} which is normal to the MIT bag surface in its
rest frame. The eigenstates of the quarks in such a spherical cavity can be
written as 
\begin{equation}
\psi _{n\kappa M}(\mbox{\boldmath{$r$}})=\left( 
\begin{array}{c}
{\frac{\displaystyle u_{n\kappa }(x)}{\displaystyle x}}{\cal Y}_{\kappa M}(%
\widehat{\mbox{\boldmath{$r$}}}) \\ 
\\ 
i{\frac{\displaystyle v_{n\kappa }(x)}{\displaystyle x}}{\cal Y}_{-\kappa M}(%
\widehat{\mbox{\boldmath{$r$}}})
\end{array}
\right) ,  \label{8}
\end{equation}
with $x=qr$ and $q$ the absolute value of the 3-momentum of the quark. Here, 
$\widehat{\mbox{\boldmath{$r$}}}$ is the unit vector pointing in the same
direction as the coordinate vector $\mbox{\boldmath{$r$}}$ for the quark.
The sub-indices $n=0,1,2,\ldots $, $\kappa =\pm 1,\pm 2,\ldots ,$ and $M$
are quantum numbers besides the flavour and color that completely specify a
single quantum state \cite{11} for a quark. The radial wave functions $%
u_{n\kappa }$ and $v_{n\kappa }$ are proportional to spherical Bessel
functions since the quarks are free particles inside the bag. The
spin-orbital angular momentum coupled wave functions, have a total angular
momentum $j=|\kappa |-{1/2}$ are ${\cal Y}_{\kappa m}$ and ${\cal Y}%
_{-\kappa m}$. The orbital angular momentum $l$ for ${\cal Y}_{\kappa m}$ is 
$j-1/2$ if $\kappa <0$ and $j+1/2$ if $\kappa >0$. Imposing Eq.(\ref{7}) on
the wave function Eq.(\ref{8}), we get 
\begin{equation}
u_{n\kappa }(qR)=v_{n\kappa }(qR),  \label{9}
\end{equation}
which can be numerically solved for each $n$ and $\kappa $ to get a discrete
set of $q_{n\kappa }$. The energy $\epsilon $ for a quark in such a state is 
\begin{equation}
\epsilon _{n\kappa }=\sqrt{q_{n\kappa }^2+m^2}.  \label{10}
\end{equation}
This has a one to one correspondence to $q_{n\kappa }$. It is expected that
the value for each $\epsilon _{n\kappa }$ has to be obtained numerically for
the MIT bag boundary condition.

The approach of summation over the microscopic energy levels to obtain the
thermodynamical potential is rather computationally expensive. For example,
the thermodynamical potential of a strangelet can be written as 
\begin{equation}
\Omega =-{\frac{n_c}\beta }\sum_{f=\{u,d,s\}}\sum_{n=0}^\infty \sum_{\kappa
=-\infty }^\infty (2j_\kappa +1)\ln \left( 1+e^{-\beta (\epsilon _{fn\kappa
}-\mu _f)}\right) +\Omega ^{\prime },  \label{11}
\end{equation}
where $\beta $ is the inverse temperature, the summation over $\kappa $
excludes $\kappa =0$, $j_\kappa $ is the total angular momentum of the state
corresponding to $\kappa $, $\mu _f$ is the chemical potential for flavor $f$
, $n_c=3$ is the degeneracy due to the color degrees of freedom of quarks,
and $\Omega ^{\prime }$ is the contribution to the thermodynamical potential
from other degrees of freedom of the system.

In most of the cases, the quantity being summed is a smooth enough function
of the quark energy on the scale that the DOS fluctuates. To study the gross
features, we can introduce the concept of DOS and replace the microscopic
summation by an integration over the quark's momentum, namely 
\begin{equation}
\sum_{\epsilon _{n\kappa }}F(\epsilon _{n\kappa })\to \int_0^\infty dk\rho
(k)F[\epsilon (k)].  \label{12}
\end{equation}
Here $\epsilon _{n\kappa }$ is the discrete energy level, $\epsilon (k)$ is
its continuous interpolation using Eq.(\ref{10}), and $\rho (k)$ is the DOS
which is determined by the total number of quark states $N(k)$. In the
numerical calculation, the $\overline{N}(k)$ that provides the best fit to $%
N(k)$ is used as a smooth representation of $N(k)$. With a smooth $\overline{%
N}(k)$, the DOS $\rho (k)$ in the above equation is then 
\begin{equation}
\rho (k)={\frac{d\overline{N}(k)}{dk}}.  \label{13}
\end{equation}

There is no guarantee that the left hand side of Eq.(\ref{12}) is the same
as its right hand side in the case studied here, since the microscopic
distribution of the energy levels does not follow a trend regular enough to
avoid the fluctuations introduced by $\overline{N}(k)$ replacing $N(k)$. In
fact, as is found in the following, the fluctuations in the DOS are large
and irregular. However, the situation can be improved by expressing Eq.(\ref
{12}) as 
\begin{equation}
\sum_{\epsilon _{n\kappa }}F(\epsilon _{n\kappa })=\int_0^\infty dk\rho
(k)F[\epsilon (k)]+<\delta F>,  \label{14}
\end{equation}
where $<\delta F>$ is by definition the difference between the left and
right hand side of Eq.(\ref{12}). It is easy to see that $<\delta F>=0$ if 
\begin{equation}
<N-\overline{N}>(k)=\sum_{k_i}\left[ N(k_i)-\overline{N}(k_i)\right] \to 0,
\label{15}
\end{equation}
where the summation is over the energy levels of quarks with a magnitude of
3-momentum in the neighbourhood of $k$ in which $F[\epsilon (k)]$ does not
change appreciably. The best we can do to find the general trend of the
problem is to extract from $N(k)$ as precisely as possible the smooth
quantity $\overline{N}(k)$ which can be attributed to the DOS . The quantity 
$<\delta F>$ is then usually small and can be ignored.

The total quark states with the momentum $\mbox{\boldmath{$q$}}\le k$ for
each flavor is counted according to 
\begin{equation}
N(k)=n_c\sum_{n=0}^\infty \sum_{\kappa =-\infty }^\infty \theta
(k-q_{n\kappa })(2j_\kappa +1),  \label{16}
\end{equation}
where $q_{n\kappa }$ is determined by Eq.(\ref{9}) and the step function $%
\theta (x)=1$ if $x>0$, and $\theta (x)=0$ if $x<0$. This number is, in
principle, a function of three variables, namely, the cutoff momentum $k$,
the mass $m$ and the bag radius $R$. Since $N(k)$ is a dimensionless number,
it is expected to be a function of dimensionless variables too. Two such
kinds of variables, namely, $m/k$ and $kR$ formed by $k$, $m$ and $R$ have
been chosen in Eq.(\ref{totNexp1}) while the other variables, for example $%
mR $, will not emerge in expressions of coefficients $A$, $B$ and $C$
because they are independent of $R$.

The numerically counted $N(k)$ is therefore fitted by using the following
smooth trial functional form ansatz 
\begin{equation}
\overline{N}(k)=A\left( \frac mk\right) \left( kR\right) ^3+B\left( \frac mk%
\right) \left( kR\right) ^2+C\left( \frac mk\right) \left( kR\right) ,
\label{ansatz}
\end{equation}
where all coefficients $A,B$ and $C$ are assumed to be independent of $kR$.

The detailed fitting is done in the following way, First, $m$ and $k$ are
kept fixed (so $m/k$ is a constant), and various corresponding $N(k)$ are
counted by varying $R$. Such a series of $N(k)$ data as a function of $R$ is
fitted using the smoothed representation Eq.(\ref{ansatz}) to obtain the
coefficients $A$, $B$ and $C$ for a fixed $m/k$. We employ ``the method of
minimum squares'' to obtain the best fit. Then the value of $m$ is changed
with a fixed $k$ and the above steps are repeated to obtain a different $A$, 
$B$ and $C$ for different $m/k$.

The consistency of the procedure is checked by calculating $N(k)$ with a
series of different cutoff momentum $k$ (25,50,100,200,400 and 800 MeV) with 
$m/k$ and $kR$ fixed. It is found that the choice of cutoff momentum will
not affect the coefficients $A(m/k)$, $B(m/k)$ and $C(m/k)$ for each $m$ and 
$k$ bin.

\section{Smoothed part of the DOS under the bag boundary condition}

The results of our best fit for smooth curves by using ``the method of
minimum squares'' can be summarized as follows.

\subsection{The volume term}

The volume term $A$ is found to be 
\begin{equation}
A\left( \frac mk\right) =7.074\times 10^{-2}g,  \label{Our-A}
\end{equation}
which is different from ${2g}/{9\pi }$ by a difference of $10^{-4}-$ $%
10^{-5} $. It means that $A({m}/{k})$ is independent of $m/k$ to a very good
approximation. This result is in good agreement with the MRT as well as the
plane wave approximation.

\subsection{The surface term}

The coefficient $B({m}/{k})$ depends on $m/k$ explicitly. The results of our
best fitted $B(m/k)$ is compared to the one from the MRT. The relative
difference between $B(m/k)$ and $B_{MRT}(m/k)$ is shown in Fig. 1. It can be
seen that such a relative difference is around $1-2$ $\%$ for most of the $%
m/k$ value except at rather small $m/k$. Thus we come to the conclusion that
the coefficient $B_{MRT}(m/k)$ describes the surface correction of the DOS
very well.

\subsection{The curvature term}

After we remove the volume term and the surface term , the remaining
numerical data for $N(k)$, which is denoted as $N_{res}^{(1)}(k)$, contains
information about the coefficient of the curvature term $C(m/k)$. The
fluctuations of $N_{res}^{(1)}(k)$ are large. To extract the linear term, an
average of the remaining data $N_{res}^{(1)}(k)$ defined by 
\begin{equation}
\overline{N}_{res}^{(1)\Delta }(kR)=\frac 1\Delta \sum_{x=kR-\delta x\Delta
/2}^{kR+\delta x\Delta /2}N_{res}^{(1)}(x),  \label{smth-def}
\end{equation}
where $\delta x$ is the difference between the neighbouring $kR$ points in
the numerical data for $N(k)$, is made first. The averaged $\overline{N}%
_{res}^{(1)\Delta }$ is fitted to a linear function by using the method of
minimum squares to get the curvature term. The resulting coefficient $C(m/k)$
is not a slow varying function of $m/k$ as is the one assumed in Eq.(\ref{6}%
). The result of our best fit for $\Delta =50$ is shown in Fig. 2 which
agrees with the multi-reflection theory at the two ends, $m/k=0$ and $m/k\to
\infty $, but is a fluctuating function of $m/k$ in between. If these
fluctuations are smoothed out, it contains a global maximum which is
different from the proposed monotonic function given in Eq.(\ref{6}). A
smooth representation of the extracted (fluctuating) function is the
following 
\begin{equation}
C(m/k)=\widetilde{C}(m/k)+\left( {\frac mk}\right) ^{1.45}{\frac g{%
3.42\left( {\frac{\displaystyle m}{\displaystyle k}}-6.5\right) ^2+100}}
\label{Our-C}
\end{equation}
with $\widetilde{C}(m/k)$ given by Eq.(\ref{6}). This is plotted as the
solid line in Fig. 2 together with the numerical data points. The dashed
line is $\widetilde{C}(m/k)$.

\subsection{Fluctuating term}

Most of the fluctuations $N(k)$ reside in the remaining part of Eq.(\ref
{totNexp1}). It is also expected that the difference between the smoothing
procedure of the MRT and the one adopted here resides in this residual term,
which is obtained from the numerical data $N(k)$ using the following
equation 
\begin{equation}
N_{res}^{(0)}=N(k)-{\frac{2g}{9\pi }}(kR)^3-B(m/k)(kR)^2-C(m/k)(kR),
\label{a2}
\end{equation}
where $B(m/k)$ and $C(m/k)$ are obtained from the procedure above described.
We use the method for $N_{res}^{(0)}$ of Eq. (\ref{smth-def}) to smooth $%
N_{res}^{(0)}$ and obtain $N_{res}^{(0)\Delta }(k)$.

The results for $\Delta =1,10,50,100$ and $m/k=0.1,100$ are shown in figure
3 and figure 4, respectively. Two trends for the numerical data are seen
after subtracting the volume, surface and curvature terms obtained by best
fitting. First the magnitude of the fluctuations decreases as $1/\sqrt{%
\Delta }$. Secondly, the magnitude of the fluctuations increases with $kR$.
The fluctuations will exist for any value of $\Delta $.

This result means that $N_{res}^{(0)}$ can not in principle be treated as
zero. According to Eq.(\ref{14}), this means that the fluctuating term $%
<\delta F>$ for any smooth function of $k$ is not zero for the MIT bag
model. Fortunately, it is not very large for a smooth function ($%
N_{res}^{(0)}/\overline{N}\approx 10^{-5}\sim 10^{-6}$) and can be dropped
in most of the applications involving the gross features of the problem.

\section{Conclusion and discussion}

The distribution of the quark states in a spherical MIT bag is calculated
numerically. By making a best fit to the data using a smooth function of the
dimensionless variables $kR$ and $m/k$, the coefficient of this kind of
power expansion is determined. It is found that the volume and surface terms
can be represented by a smooth function of the above mentioned variables in
agreement with the MRT. The curvature term which contains fluctuations is
extracted from the numerical data and represented by a smooth function, as
given by Eq.(\ref{Our-C}). The fluctuations remain after smoothing, but can
be ignored in most problems.

The same treatment can also be applied to the non-relativistic case: For
simplicity we only give the final result here. The DOS for non-relativistic
particles confined in a spherical cavity of radius $R$ with boundary
condition 
\begin{equation}
u_\xi (kR)=0  \label{a3}
\end{equation}
can be found and the best smooth fit to the numerical result is 
\begin{equation}
N(k)={\frac{2g}{9\pi }}(kR)^3-0.1289g(kR)^2+1.0144g(kR)-60.319g.  \label{a4}
\end{equation}

A non-spherical bag is needed to represent a droplet at high angular
momentum, which is expected to be produced in relatively large number in a
high energy relativistic heavy ion collision. This is not studied in this
work. A representation and numerical study of these deformed bags should be
carried out in the future.

\section*{Acknowledgment}

This work is supported in part by the National Natural Science Foundation of
China under contract Nos. 19975010, 19947001 and 19875009, respectively, and
research funds from the Ministry of Education of China.

\section{Figure Captions}

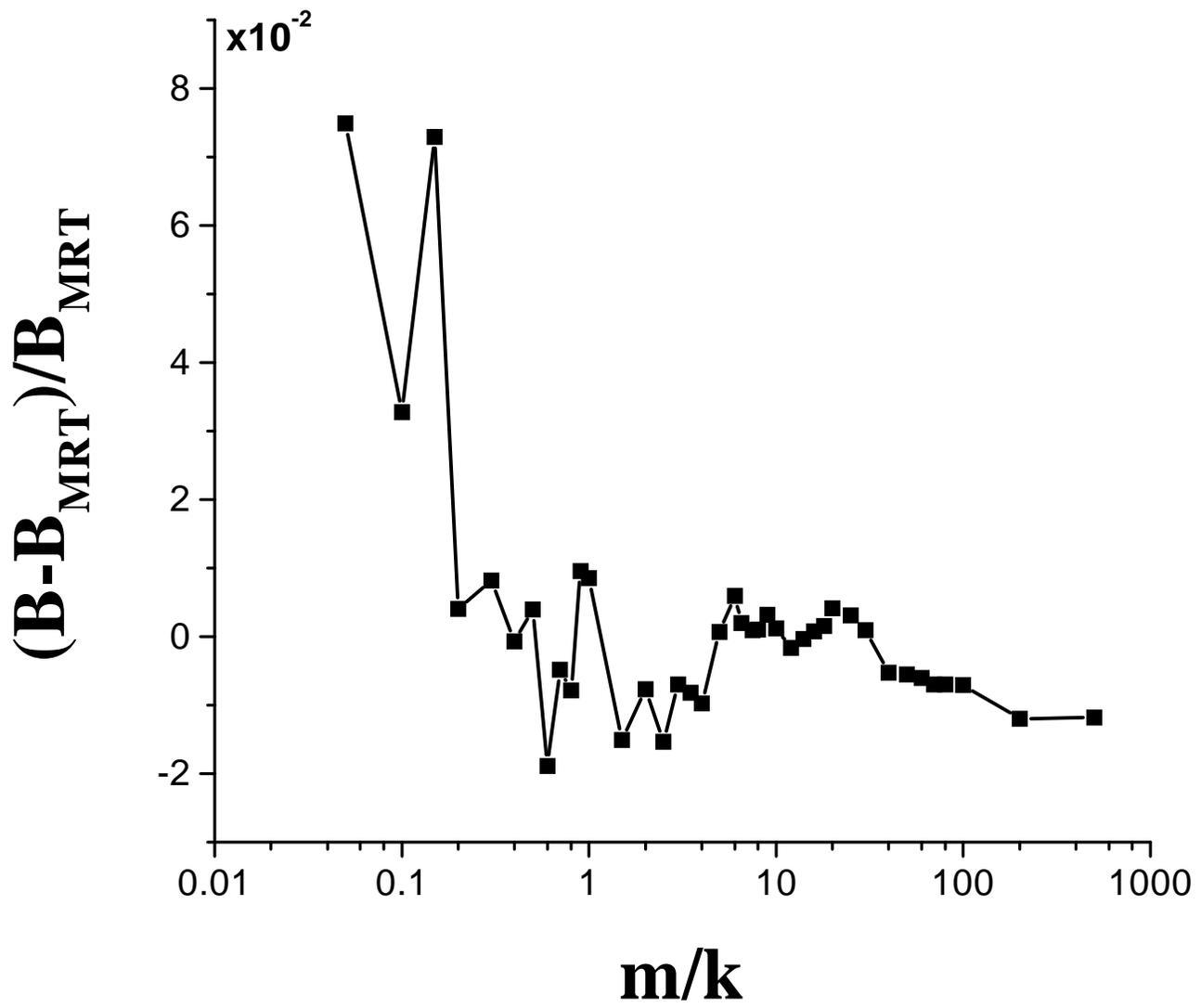
FIG.1. The difference of the coefficient $B$ as a smooth function of $m/k$
between the one best fitted to the numerical counted value and the one
calculated analytically. Their difference draw in line is almost a constant
with a variation less or equals to $1\%$ of its own value.

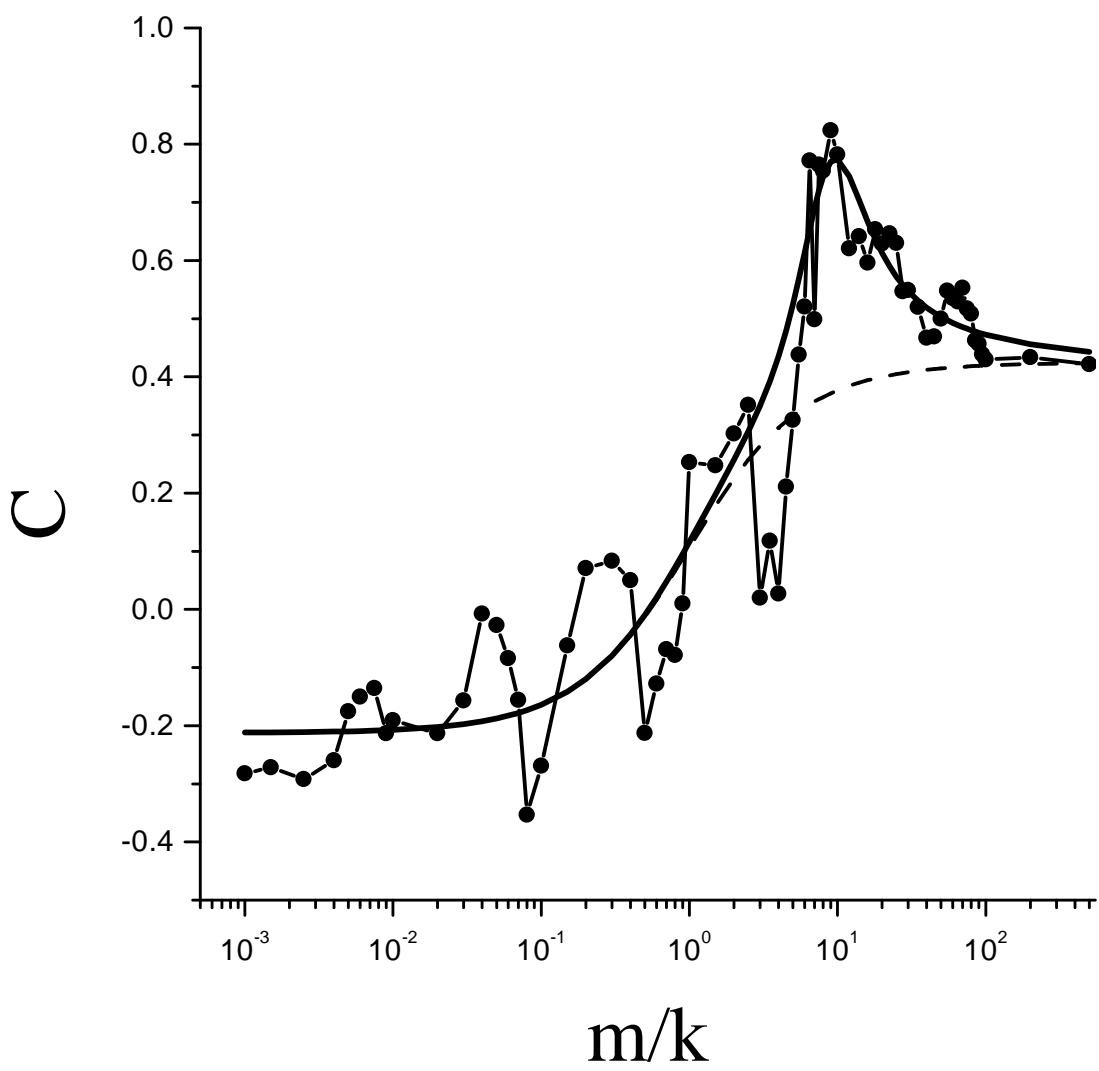
FIG.2. The coefficient $C$ extracted from $\overline{N}_{res}^{(1)\Delta }.$
The solid line is a smooth representation of it. The dashed line represents $%
\widetilde{C}(m/k)$.

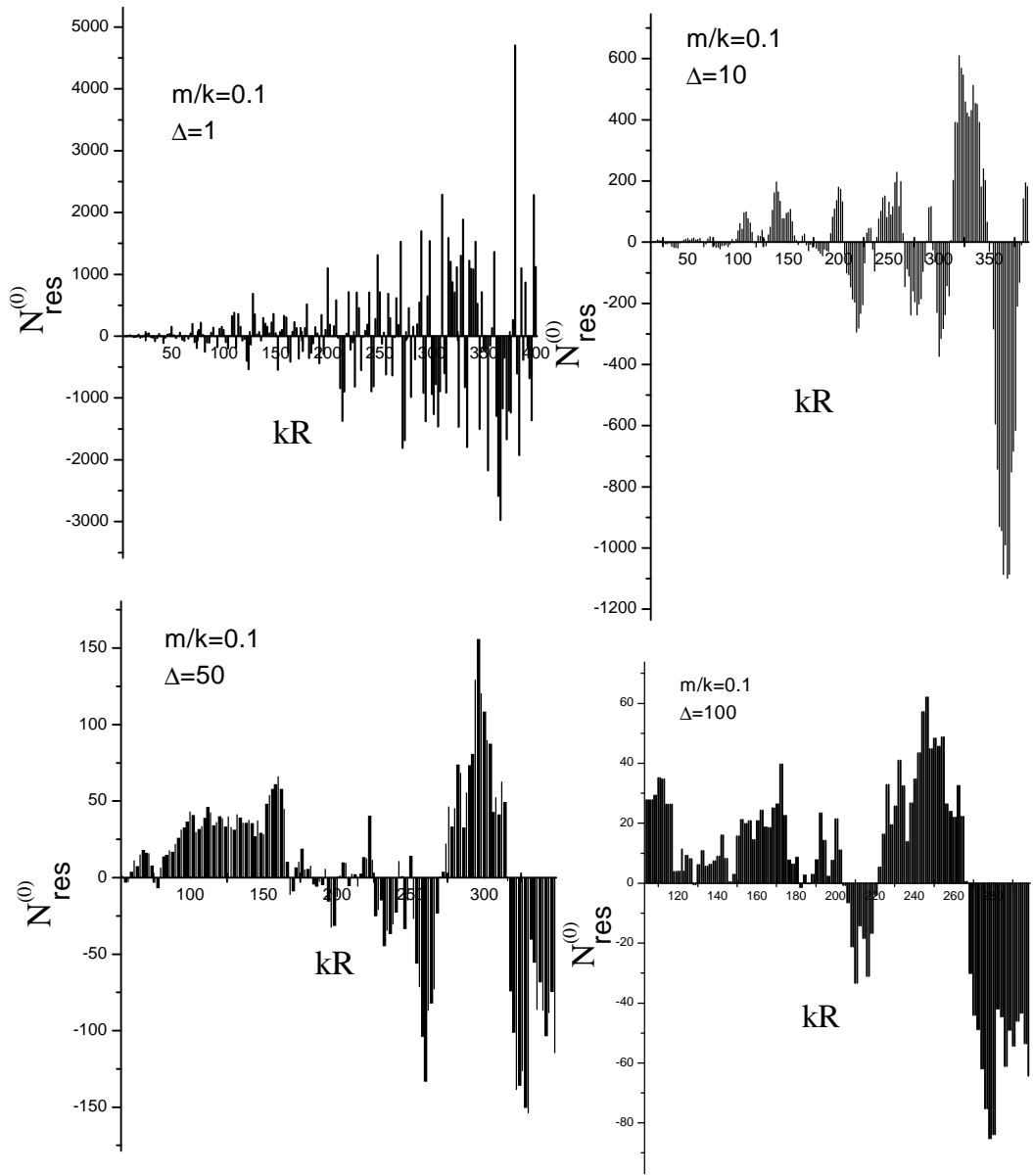
FIG.3. The behaviour of the fluctuating residual term $N_{res}^{(0)}$ under
the smoothing for $m/k=0.1$.

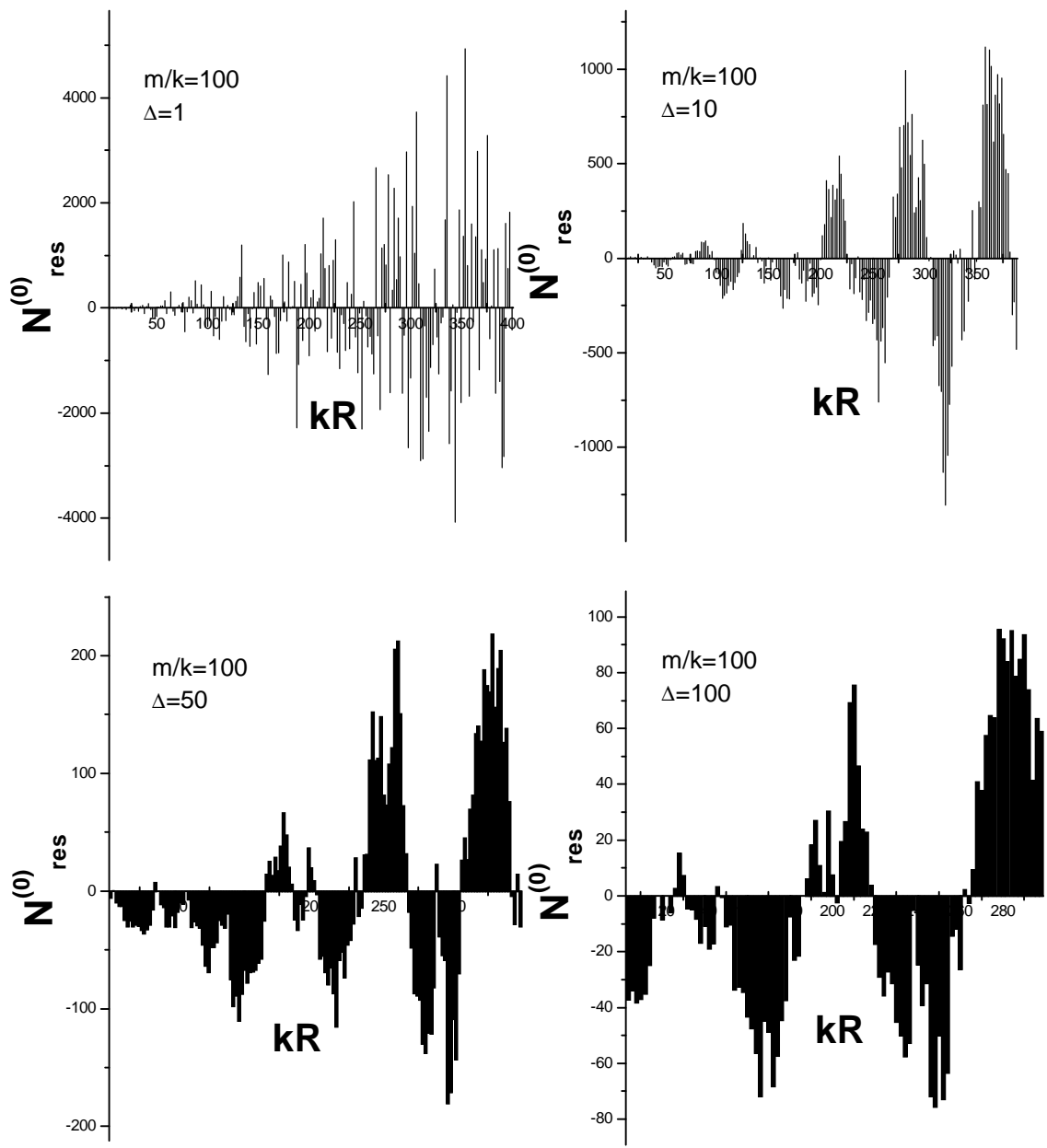
FIG.4. The behaviour of the fluctuating residual term $N_{res}^{(0)}$ under
the smoothing for $m/k=100$.

\setcounter{page}{0}

\begin{figure}[tbp]
\end{figure}

\begin{figure}[tbp]
\end{figure}

\begin{figure}[tbp]
\end{figure}


\begin{references}
\bibitem{1}  Thomas A W 1984 Advances in Nuclear Physics, vol.{\bf \ 13 }ed.
Negele J W and Vagt E ( NewYork: Plenum ) and refs. herein.

\bibitem{2}  Lee T D 1981 Particle Physics and Introduction to Field Theory,
Harwood Acad. Pub

\bibitem{3}  Guichon P AM 1988 Phys. Lett. {\bf B200,} 235; Song H Q and Su
R K 1996 J. Phys. G: Nucl.Part.Phys. {\bf 22, }1025

\bibitem{3-1}  Allen E 1975 Phys. Lett. {\bf B57}, 263

\bibitem{4}  Wang P, Su R K, Song H Q and Zhang L L 1999 Nucl. Phys. {\bf %
A653,} 166

\bibitem{5}  Berger M and Jaffe R L 1987 Phys. Rev. {\bf D35}, 213

\bibitem{6}  Mardor J and Svetitsky B 1991 Phys. Rev. {\bf D44}, 878

\bibitem{7}  Lee K S and Heinz U 1993 Phys. Rev. {\bf D47}, 2068

\bibitem{8}  Greiner C, Koch P and St\"{o}cker H 1987 Phys. Rev. Lett. {\bf %
58,} 1825

\bibitem{9}  Farhi E and Jaffe R L 1984 Phys. Rev. {\bf D30}, 2379

\bibitem{10}  Jaqaman H, Mekjian A Z and Zamick L 1984 Phys. Rev. {\bf C29,}
2067

\bibitem{11}  Balian R and Block C 1970 Ann. Phys. {\bf 60}, 401

\bibitem{12}  Hansson T H and Jaffe R L 1987 Phys. Rev. {\bf D35}, 213

\bibitem{13}  Madsen J 1994 Phys. Rev. {\bf D50}, 3328

\bibitem{14}  Madsen J 1993 Phys. Rev. {\bf D47}, 5156

\bibitem{15}  Gilson E P and Jaffe R L 1993 Phys. Rev. Lett. {\bf 71,} 332

\bibitem{16}  Chodos A et al. 1974 Phys. Rev. {\bf D7}, 3471
\end{references}
\end{document}